\documentclass[prd,twocolumn,showpacs,preprintnumbers,amsmath,amssymb,superscriptaddress,floatfix,nofootinbib]{revtex4-2}

\usepackage{graphicx}
\usepackage{bm}
\usepackage{amsmath}
\usepackage{amsfonts}
\usepackage{amssymb}
\usepackage{color}
\usepackage{multirow}
\usepackage{subfigure}
\usepackage{txfonts}
\usepackage[colorlinks, citecolor=blue,anchorcolor=red,menucolor=red, linkcolor=red,filecolor=red,runcolor=red,urlcolor=blue,frenchlinks=red]{hyperref}

\begin{document}

\title{Search for the low-lying excited baryon $\Sigma^*(1/2^-)$ through process $\Lambda^+_c \to  \Lambda K^0 \pi^+$}
 
\author{Sheng-Chao Zhang}
\affiliation{School of Physics, Zhengzhou University, Zhengzhou 450001, China}
\affiliation{School of Physics and Electronics, Henan University, Kaifeng 475004, China}
\author{Wen-Tao Lyu}
\affiliation{School of Physics, Zhengzhou University, Zhengzhou 450001, China}
\affiliation{Departamento de Física Teórica and IFIC, Centro Mixto Universidad de Valencia-CSIC Institutos de Investigación de Paterna, 46071 Valencia, Spain}

\author{Guan-Ying Wang\footnote{wangguanying@henu.edu.cn}}
\affiliation{School of Physics and Electronics, Henan University, Kaifeng 475004, China}

\author{Bo-Qiang Ma\footnote{mabq@pku.edu.cn}}
\affiliation{School of Physics, Zhengzhou University, Zhengzhou 450001, China}

\author{En Wang\footnote{wangen@zzu.edu.cn}}
\affiliation{School of Physics, Zhengzhou University, Zhengzhou 450001, China}

\date{\today}

\begin{abstract}
Motivated by recent BESIII measurements of the singly Cabibbo-suppressed processes $\Lambda^+_c \to \Lambda K^+ \pi^0$ and $\Lambda^+_c \to \Lambda K_S^0 \pi^+$, we investigate the process $\Lambda^+_c \to \Lambda K^0 \pi^+$ by taking into account the contribution from the low-lying excited baryon $\Sigma^*(1/2^-)$,  dynamically generated via the $S$-wave pseudoscalar meson-octet baryon interaction, as well as from the intermediate resonances $K^*(892)$ and $N(1535)$. 
Our model successfully reproduces the BESIII $\pi^+K^0$ invariant mass distribution, and predicts a distinct cusp structure around 1.43~GeV in the $\pi^+\Lambda$ invariant mass distribution, which is associated with the predicted $\Sigma^*(1/2^-)$. Future high-precise measurements of this process at BESIII, Belle~II, and the proposed Super Tau-Charm Facility experiments will be crucial for testing the existence of $\Sigma^*(1/2^-)$ and advancing our understanding of the light baryon spectrum.

\end{abstract}

\maketitle

\section{Introduction}

 Since the observation of the charmonium-like state $X(3872)$ by the Belle Collaboration in 2003~\cite{Belle:2003nnu}, 
 numerous exotic state candidates have been reported. 
 Understanding their properties is essential for probing the non-perturbative nature of quantum chromodynamics (QCD)~\cite{Chen:2016spr,Chen:2022asf,Guo:2017jvc,Oset:2016lyh,Liu:2024uxn}.
 In the light baryons sector, one of the outstanding puzzles is the ``mass reverse problem", where the $N(1535)$ state with the spin-parity quantum numbers of $J^P=1/2^-$ is expected to be lighter than the radially excited  $N(1440)$ state with $J^P=1/2^+$, yet experimental observations show the opposite~\cite{ParticleDataGroup:2024cfk}. Additionally,  the low-lying excited baryon $\Lambda(1405)$ with $J^P=1/2^-$ is dynamically generated from the $\bar{K}N$ interaction in the unitary chiral approach~\cite{Oset:1997it,Jido:2003cb}, and the low-lying excited $\Sigma^*(1/2^-)$ remains experimentally and theoretically unestablished~\cite{ParticleDataGroup:2024cfk}. In the Review of Particle Physics (RPP)~\cite{ParticleDataGroup:2024cfk}, although the state $\Sigma(1620)$ with $J^P=1/2^-$ is listed, it carries only one star rating (poor evidence), and is  omitted from summary table, indicating that its existence requires confirmation.

Establishing the state $\Sigma^*(1/2^-)$ is crucial to deepening our understanding of low-lying excited baryons~\cite{Crede:2013kia,Klempt:2009pi,Oset:2016lyh,Wang:2024jyk,Zou:2007mk}. Theoretical studies within the chiral unitary approach predict a $\Sigma^*(1/2^-)$ with a mass near the $\bar{K}N$ threshold from $S$-wave meson-baryon interactions in the strangeness $S = -1$ sector~\cite{Oset:2001cn,Oller:2000fj,Oset:1997it,Khemchandani:2018amu,Kamiya:2016jqc,Jido:2003cb,Oller:2006jw,Garcia-Recio:2002yxy,Lutz:2001yb}.  A refined analysis of the CLAS data on the process $\gamma p \to K\Sigma\pi$ suggests a $\Sigma^*(1/2^-)$ peak around 1430~MeV~\cite{CLAS:2013rjt,Roca:2013cca}. Meanwhile, studies of the process $K^-p\to \Lambda\pi^+\pi^-$ hint at a possible $\Sigma^*(1/2^-)$ resonance near 1380~MeV~\cite{Wu:2009nw,Wu:2009tu}. 
It is also suggested in Ref.~\cite{Gao:2010hy} that the $\Sigma(1380)$ state plays a role in the $K\Sigma^*$ photoproduction  within the framework of the effective Lagrangian approach.

Furthermore, $\Sigma^*(1/2^-)$ has been investigated in $\Lambda_c^+$ four-body decays $\Lambda^+_c\to \Sigma^0\pi^+\pi^0\pi^0$, $\Lambda^+_c\to \Sigma^+\pi^+\pi^0\pi^-$ and $\Lambda^+_c\to \Lambda\pi^+\pi^+\pi^-$ via triangle singularity mechanisms~\cite{Dai:2018hqb,Xie:2018gbi,Li:2025yad}.
In addition, it is also suggested to search for the $\Sigma^*(1/2^-)$ in the processes of $\Lambda_c^+\to\eta\Lambda\pi^+$~\cite{Xie:2017xwx,Lyu:2024qgc}, $\Xi_c^+\to\Lambda\bar{K}^0\pi^+$~\cite{Li:2025exm}, $\chi_{c0}(1P)\to\bar{\Lambda}\Sigma\pi$~\cite{Liu:2017hdx}, $\chi_{c0}(1P)\to\bar{\Sigma}\Sigma\pi$~\cite{Wang:2015qta}, and $\gamma N\to K\Sigma^{*}$~\cite{Lyu:2023oqn,Kim:2021wov}.
One can find a recent review on the low-lying excited baryon $\Sigma^*(1/2^-)$ in Ref.~\cite{Wang:2024jyk}.

 Following the suggestion of Ref.~\cite{Lyu:2024qgc}, the BESIII Collaboration analysed the process $\Lambda_c^+\to\eta\pi^+\Lambda$ and reported the evidence of $\Sigma(1380)$ $(J^P=1/2^-)$ with a statistical significance exceeding $3\sigma$~\cite{BESIII:2024mbf}. 
However, it was later shown that the BESIII measurements of this process could be well described without including the $\Sigma(1380)$ $(J^P=1/2^-)$~\cite{Duan:2024czu}.

It is notable that the Belle Collaboration has measured the process  $\Lambda_c^+\to \Lambda\pi^+\pi^+\pi^-$, and  the $\Lambda\pi^+$ and $\Lambda\pi^-$ invariant mass distributions revealed clear cusp structure, which could be interpreted as a resonance with mass of $(1434.3\pm0.6\pm 0.9)$~MeV and width of $(11.5\pm2.8\pm 5.3)$~MeV for the $\Lambda\pi^+$ combination, or with mass of $(1438.5\pm0.9\pm 2.5)$~MeV and width of $(33.0\pm 7.5\pm 23.6)$~MeV for the $\Lambda\pi^-$ combination~\cite{Belle:2022ywa}. Its average mass is consistent with the predictions for the $\Sigma^*(1/2^-)$, dynamically generated from the $S$-wave pseudoscalar meson-octet baryon interaction within the chiral unitary approach~\cite{Oset:1997it,Oller:2000fj,Roca:2013cca,Guo:2012vv,Jido:2003cb}. Thus, confirming the $\Sigma^*(1/2^-)$ in other processes remains imperative.

The non-leptonic weak decays of charmed baryons are an important laboratory for studying light hadrons, due to their large phase space and significant final-state interactions~\cite{Oset:2016lyh,Miyahara:2015cja,Li:2026lbo,Wang:2022nac,Wang:2020pem,Zeng:2020och,Feng:2020jvp,Li:2024rqb,Zhang:2024jby,Wang:2024jyk,Li:2025gvo}. Recently, the BESIII Collaboration has reported the first observation of the singly Cabibbo-suppressed decay $\Lambda^+_c\to\Lambda K^+\pi^0$ with a significance of 5.7$\sigma$ and obtained the branching fraction $\mathcal{B}(\Lambda_c^+ \to \Lambda K^+\pi^0)=(1.49\pm0.27_{\text{stat}}\pm0.05_{\text{syst}}\pm0.08_{\text{ref}})\times 10^{-3}$~\cite{BESIII:2023sdr}. Subsequently, using integrated luminosity  4.5~fb$^{-1}$ of $e^+e^-$ collision data, BESIII also measured the branching fractions $\mathcal{B}(\Lambda_{c}^{+} \to \Lambda K_S^{0}\pi^{+}) = (1.73 \pm 0.27 \pm 0.10)\times 10^{-3}$~\cite{BESIII:2024xny}, and presented only the $K_S^0\pi^+$ invariant mass distribution. Given that $\Sigma^*(1/2^-)$ is predicted to couple strongly to the $\pi\Lambda$ channel~\cite{CLAS:2013rjt,Roca:2013cca},  supported by the Belle measurements of the $\Lambda_c^+\to \Lambda\pi^+\pi^+\pi^-$~\cite{Belle:2022ywa}, it is expected that the $\Sigma^*(1/2^-)$ could play an important role in the process $\Lambda_{c}^{+} \to \Lambda K^{0}\pi^{+}$. Therefore, we propose to search for the low-lying excited baryon $\Sigma^*(1/2^-)$ via the $\Lambda\pi$ mass distribution in $\Lambda_c^+\to\Lambda K^0\pi^+$ decay. 

The BESIII measurements on the $K_S^0\pi^+$ invariant mass distribution of the process $\Lambda_{c}^{+} \to \Lambda K_S^{0}\pi^{+}$  show a peak structure around 0.9~GeV, associated with the $K^*(892)$. Moreover, the $N(1535)$ with $J^P=1/2^-$ is known to couple strongly to the $K\Lambda$ channel~\cite{Liu:2005pm}. In this work, we will investigate the singly Cabibbo-suppressed process $\Lambda^+_c \to  \Lambda K^0 \pi^+ $ by incorporating $S$-wave pseudoscalar meson-octet baryon interaction  within the chiral unitary approach, which  dynamically generates the resonances $\Sigma^*(1/2^-)$ and $N(1535)$,  along with the intermediate vector meson $K^*(892)$. Our analysis aims to motivate more precise experimental investigations.

This article is organized as follows. In Sec.~\ref{Sec:Formalism}, we present the theoretical framework for the $\Lambda^+_c \to \Lambda K^0 \pi^+$ decay. Numerical results and discussions are presented in Sec.~\ref{Sec:Results}. Finally, a short summary and an outlook are provided in Sec.~\ref{Sec:summary}.

\section{Theoretical Formalism} \label{Sec:Formalism}

\begin{figure*}[htbp]
  \centering
  \subfigure[]{%
    \includegraphics[scale=0.8]{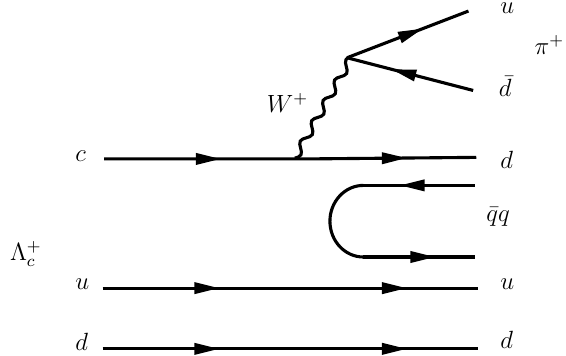}
  }
  \hfill 
  \subfigure[]{%
    \includegraphics[scale=0.8]{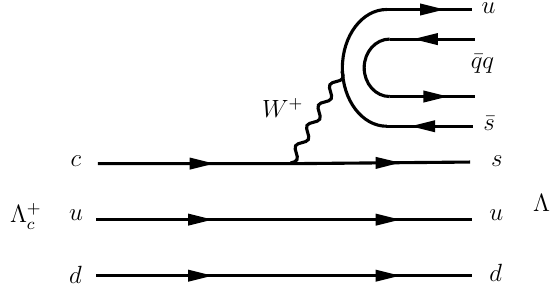}
  }
  \par 

  \subfigure[]{%
    \includegraphics[scale=0.8]{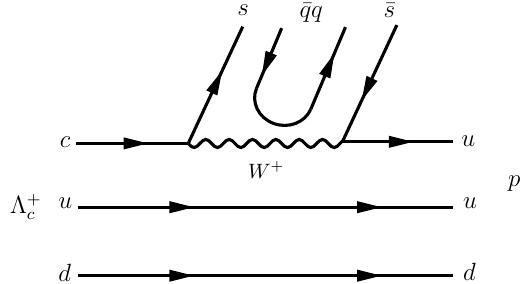}
  }
  \hfill
  \subfigure[]{%
    \includegraphics[scale=0.8]{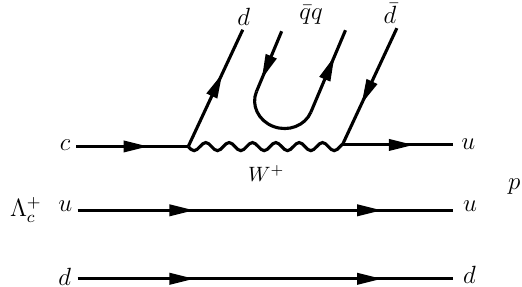}
  }

  \caption{Feynman diagrams at the quark level. 
  (a) $\Lambda_{c}^{+}\rightarrow \pi^{+}d(\bar{u}u+\bar{d}d+\bar{s}s)ud$, (b) $\Lambda_{c}^{+}\rightarrow u(\bar{u}u+\bar{d}d+\bar{s}s)\bar{s}\Lambda$, 
  (c) $\Lambda_{c}^{+}\rightarrow s(\bar{u}u+\bar{d}d+\bar{s}s)\bar{s}p$, and (d) $\Lambda_{c}^{+}\rightarrow d(\bar{u}u+\bar{d}d+\bar{s}s)\bar{d}p$.}
\label{Fig:wei}
\end{figure*}

We begin with the dominant color-favored $ W^{+}$ external emission mechanism and the color-suppressed $ W^{+}$ internal emission mechanism for the decay $\Lambda^+_c \to \Lambda K^0 \pi^+$, depicted in Fig.~\ref{Fig:wei}. For Fig.~\ref{Fig:wei}(a), the $c$ quark of the initial $\Lambda_c^+$ weakly decays into a $W^+$ boson and a $d$ quark, then the ${W^{+}}$ boson decays into a $u\bar{d}$ quark pair. The $u\bar{d}$ quark pair from the ${W}^+$ boson will hadronize into $\pi^{+}$, while the $d$ quark and the $ud$ quark pair of the initial $\Lambda_c^+$, together with the quark pair $\bar{q}q=\bar{u}u+\bar{d}d+\bar{s}s$ created from the vacuum with the quantum numbers $J^{PC}=0^{++}$, will hadronize into a baryon-meson pair, which could be expressed as,

\begin{align}
\Lambda_c^+ &\Rightarrow \frac{1}{\sqrt{2}}c(ud-du)\chi_{MA} \nonumber\\
&\Rightarrow \frac{1}{\sqrt{2}}W^+d(ud-du)\chi_{MA} \nonumber\\
&\Rightarrow \frac{1}{\sqrt{2}}u\bar{d} d(\bar{u}u+\bar{d}d+\bar{s}s)(ud-du)\chi_{MA} \nonumber \\
&\Rightarrow \frac{1}{\sqrt{2}}\pi^+\sum M_{2i} q_i(ud-du) \chi_{MA},  
\end{align}
where $\chi_{MA}$ denotes the mixed antisymmetric spin wave function of the $(ud-du)$ diquark, and $M$ is the matrix of the pseudoscalar mesons,
\begin{eqnarray}
	& M & = \left(\begin{array}{cccc}u\bar{u}&u\bar{d}&u\bar{s}\\
		d\bar{u}&d\bar{d}&d\bar{s}\\
		s\bar{u}&s\bar{d}&s\bar{s}
		\end{array}
		\right) \nonumber\\
	&&= \left(\begin{matrix} \frac{\eta}{\sqrt{3}}+ \frac{{\pi}^0}{\sqrt{2}}+ \frac{{\eta^\prime}}{\sqrt{6}} & \pi^+ & K^+  \\
		\pi^-  &   \frac{\eta}{\sqrt{3}}- \frac{{\pi}^0}{\sqrt{2}}+ \frac{{\eta^\prime}}{\sqrt{6}}  &  K^0 \\
		K^-  &  \bar{K}^{0}   &    -\frac{\eta}{\sqrt{3}}+ \frac{{\sqrt{6}\eta^\prime}}{3}
	\end{matrix}
	\right),
\end{eqnarray} 
where we have considered the approximate $\eta-\eta'$ mixing~\cite{Bramon:1992kr,Lyu:2023ppb,Li:2025gvo}, and the channels including the $\eta^{\prime}$ component are ignored since the $\eta^{\prime}$ has a large mass. 

In the next step, we obtain the components of the pseudoscalar meson and octet baryon as,
\begin{align}
&\quad \frac{1}{\sqrt{2}}\pi^+\sum M_{2_{i}} q_{i}(ud-du) \chi_{MA}
 \nonumber\\
&=\pi^+ \frac{1}{\sqrt{2}} \left(\pi^-p-\frac1{\sqrt{2}}\pi^0n+\frac1{\sqrt{3}}\eta n+\sqrt{\frac23}K^0\Lambda \right) \label{eq:MB}
\nonumber\\
&=\pi^+ \left(\frac{1}{\sqrt{2}}\pi^-p-\frac{1}{2}\pi^0n+\frac1{\sqrt{6}}\eta n+\frac1{\sqrt{3}}K^0\Lambda \right). 
\end{align} 
The baryon octet wave function is given by $\psi = \frac{1}{\sqrt{2}} \bigl( \phi_{MS} \chi_{MS} + \phi_{MA} \chi_{MA} \bigr)$, where $\phi_{MS}\chi_{MS}$ and $\phi_{MA}\chi_{MA}$ denote the mixed-symmetric and mixed-antisymmetric flavor-spin wave functions. As reported in Refs.~\cite{Pavao:2017cpt,Miyahara:2016yyh}, the wave function $\phi_{MA}$ of the baryon is defined as follows, 
\begin{gather}
		p=	\frac{u(ud-du)}{\sqrt{2}}, \\
		n=\frac{d(ud-du)}{\sqrt{2}}, \\
		\Lambda=\frac{u(ds-sd)+d(su-us)-2s(ud-du)}{2\sqrt{3}}.
\end{gather} 

Alternatively, one could hadronize the $u \bar{d}$ pair of Fig.~\ref{Fig:wei}(a) with the quark pair $q\bar{q}$ as follows:
\begin{eqnarray}
	u\bar{d} \to\left(\frac{\pi^0}{\sqrt{2}}+\frac{\eta}{\sqrt{3}}\right)\pi^++\pi^+\left(-\frac{\pi^0}{\sqrt{2}}+\frac{\eta}{\sqrt{3}}\right)+K^+\bar{K}^0. \nonumber \\
\end{eqnarray}
A subtle point here, as discussed in detail in Refs.~\cite{Duan:2024czu,Duan:2024okk}, is that the $\eta\pi^+$ and $\pi^+\eta$ terms do not add up but rather cancel each other out. This cancellation arises from the $[P,\partial_\mu P]W^\mu$ structure of the $WPP$ vertex~\cite{Gasser:1983yg,Scherer:2002tk,Ren:2015bsa,Sun:2015uva}, where $P$ denotes the pseudoscalar meson matrix. 
Similarly, the $K^+\partial_0\bar{K}^0-\bar{K}^0\partial_0 K^+$ structure in the $K^+\bar{K}^0$ term also yields no net contribution~\cite{Duan:2024czu,Sun:2015uva}. Therefore, we do not consider these hadronization channels.

Similarly, we can obtain the expression for Fig.~\ref{Fig:wei}(b), 
\begin{align}
\Lambda_c^+ &\Rightarrow \frac{1}{\sqrt{2}}c(ud-du)\chi_{MA} \nonumber\\
&\Rightarrow \frac{1}{\sqrt{2}}u(\bar{q}q)\bar{s}s(ud-du)\chi_{MA} \label{eq:7b}\nonumber\\
&\Rightarrow \sum M_{1i}M_{i3} \frac{1}{\sqrt{2}}s (ud-du) \chi_{MA} \nonumber \\ 
&=\left(\frac{1}{\sqrt{6}}\pi^0K^+ +\frac{1}{\sqrt{3}}\pi^+K^0\right)\Lambda.  
\end{align}
Among them, $\frac{1}{\sqrt{3}}\pi^+K^0\Lambda$ contributes to the final state.

Fig.~\ref{Fig:wei}(c) can be expressed as, 
\begin{align}
\Lambda_c^+ &\Rightarrow \frac{1}{\sqrt{2}}c(ud-du)\chi_{\rm MA} \nonumber\\
&\Rightarrow \frac{1}{\sqrt{2}}s(\bar{q}q)\bar{s}u(ud-du)\chi_{\rm MA} \label{eq:8c} \nonumber\\ 
&\Rightarrow \sum M_{3i}M_{i3} \frac{1}{\sqrt{2}}u(ud-du) \chi_{\rm MA} \nonumber \\
&= \left(\frac{1}{\sqrt{2}}K^-K^+ +\frac{1}{\sqrt{2}}\bar{K}^0K^0+\frac{1}{3\sqrt{2}}\eta\eta\right)p. 
\end{align}

The expression for Fig.~\ref{Fig:wei}(d) reads, 
\begin{align}
\Lambda_c^+ &\Rightarrow \frac{1}{\sqrt{2}}c(ud-du)\chi_{\rm MA} \nonumber\\
&\Rightarrow \frac{1}{\sqrt{2}}d(\bar{q}q)\bar{d}u(ud-du)\chi_{\rm MA} \label{eq:9d} \nonumber\\ 
&\Rightarrow \sum M_{2i}M_{i2} \frac{1}{\sqrt{2}}u(ud-du) \chi_{\rm MA} \nonumber \\
&= \left(\frac{1}{\sqrt{2}}\pi^- \pi^+ + \frac{1}{3\sqrt{2}}\eta\eta -\frac{1}{\sqrt{3}}\eta \pi^0 \right. \nonumber\\
&\quad \left. +\frac{1}{2\sqrt{2}}\pi^0\pi^0+\frac{1}{\sqrt{2}}K^0\bar{K}^0\right)p.  
\end{align}

For Eqs.~(\ref{eq:8c}) and (\ref{eq:9d}), only the $K^0\bar{K}^0 p$ term from Eqs.~(\ref{eq:8c}) and (\ref{eq:9d}) and the $\pi^- \pi^+ p$ term from Eq.~(\ref{eq:9d}) contribute to the final state of the $\Lambda_c^+ \to \Lambda K^0 \pi^+$ decay via the transitions of $\bar{K}^0p\to \pi^+\Lambda $ and $\pi^+p\to K^0\Lambda$.

\begin{figure}[htbp]
		\centering
	\subfigure[]{
		\includegraphics[scale=0.8]{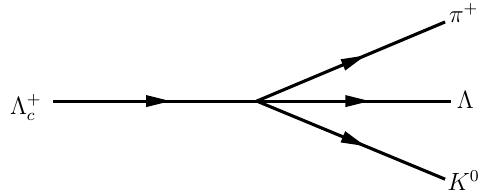}
	}
	
	\subfigure[]{
		\includegraphics[scale=0.8]{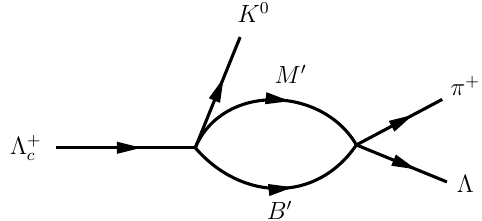}
	}
	
	\subfigure[]{
		\includegraphics[scale=0.8]{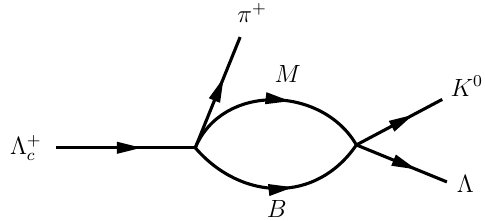}
	}
	\caption{Mechanisms for the process $\Lambda^+_c \to \Lambda K^0 \pi^+$. (a) tree diagram, (b) the $\pi^+\Lambda$ final interaction,} and (c) the $K^0\Lambda$ final interaction. \label{fig:hadron_tree_loop}
\end{figure}

Thus, the process of $\Lambda^+_c \to \Lambda K^0 \pi^+$ decay could proceed via the tree level diagram [Fig.~\ref{fig:hadron_tree_loop}(a)], the $\pi^+\Lambda$ final interaction dynamically generates the $\Sigma^*(1/2^-)$ state [Fig.~\ref{fig:hadron_tree_loop}(b)], and the $K^0\Lambda$ final interaction dynamically generates the  $N(1535)$ [Fig.~\ref{fig:hadron_tree_loop}(c)].
 
\subsection{Contribution from $\Sigma^*(1/2^-)$} \label{Subsec:A}


Taking into account that $\Sigma^*(1/2^-)$ could be generated via meson-baryon interaction, as done in Refs.~\cite{Dai:2018hqb,Xie:2018gbi,Roca:2013cca}, we can write down the decay amplitude for Fig.~\ref{fig:hadron_tree_loop}(b) as follows,
\begin{align}
\mathcal{T}^{\Sigma^*(1/2^-)} &= V_P \left[ 
    h_{K^0 \Lambda} G_{\pi^+ \Lambda} t_{\pi^+ \Lambda \to \pi^+ \Lambda} \right.\nonumber \\
    &\quad + h_{\pi^+K^0} G_{\pi^+ \Lambda} t_{\pi^+ \Lambda \to \pi^+ \Lambda}  \nonumber\\
    &\quad + \frac{2}{C} h_{\bar{K}^0K^0} G_{\bar{K}^0 p} t_{\bar{K}^0 p \to \pi^+ \Lambda}
    \left. \right],
\label{eq:T-sigma}
\end{align}
where $h_{K^0\Lambda}=\frac{1}{\sqrt{3}}$, $h_{\pi^+K^0}=\frac{1}{\sqrt{3}}$, and $h_{\bar{K}^0K^0}=\frac{1}{\sqrt{2}}$, and these values can be obtained from Eq.~(\ref{eq:MB}) and Eqs.~(\ref{eq:7b}-\ref{eq:9d}). $C$ is the color factor, and we take $C=3$ herein. The transition amplitudes $t_{\pi^+ \Lambda \to \pi^+ \Lambda}$ and $t_{\bar{K}^0 p \to \pi^+ \Lambda}$ can be obtained by solving the Bethe-Salpeter equation as follows, 
\begin{eqnarray}
T=[1-VG]^{-1}V,
 \label{Eq:BS}
 \end{eqnarray}
where the transition potential $V_{ij}$ represents a $3\times3$ matrix for the interaction kernal with three coupling channels $\bar{K}N$, $\pi\Sigma$, and $\pi\Lambda$. This potential can be taken from Ref.~\cite{Roca:2013av}, 
\begin{eqnarray}
\begin{aligned}
V_{ij}& =-\frac{C_{ij}}{4f^2}\left(2\sqrt{s}-M_i-M_j\right)  \\
&\times\left(\frac{M_i+E_i}{2M_i}\right)^{1/2}\left(\frac{M_j+E_j}{2M_j}\right)^{1/2},
 \label{eq:Vij}
\end{aligned}
 \end{eqnarray}
where $f$ is the average meson decay constant, with $f=1.123f_{\pi}$ and $f_{\pi}=92.4$ MeV.
The values of coefficients $C_{ij}$ are presented in Table~\ref{tab:Cij}, where each coefficient of the potentials of the unitary amplitudes is multiplied by one real parameter $\alpha^1_{ij}$ to give rise to the resonance $\Sigma^*(1/2^-)$~\cite{Roca:2013cca}, and we tabulate these parameters in Table~{\ref{tab:1}}.
 \begin{table}[htbp]
	\begin{center}	
	\caption{Coefficients $C_{ij}$ of the potentials~\cite{Roca:2013cca}.} 	\label{tab:Cij}
	\begin{tabular}{cccc}
		\hline\hline  
		            \qquad\quad& $\bar{K}N$  \qquad\quad& $\pi\Sigma$  \qquad\quad& $\pi\Lambda$ \\ \hline
		$\bar{K}N$     \qquad\quad&$\alpha_{11}^1$           \qquad\quad&$-~\alpha_{12}^1$          \qquad\quad&$-\sqrt{\frac{3}{2}}\alpha_{13}^1$      \\
		$\pi\Sigma$    \qquad\quad&$-~\alpha_{12}^1$         \qquad\quad&2$\alpha_{22}^1$         \qquad\quad&0      \\
		$\pi\Lambda$  \qquad\quad& $-\sqrt{\frac{3}{2}}\alpha_{13}^1$         \qquad\quad&0            \qquad\quad&0          \\
		\hline\hline
	\end{tabular}
	\end{center}
\end{table}

The $G$ in Eqs.~(\ref{eq:T-sigma}) and (\ref{Eq:BS}) is the loop function of the meson-baryon system~\cite{Inoue:2001ip},
\begin{equation}\label{loop function}
G_i=i\int\frac{d^4q}{(2\pi)^4}\frac{2M_i}{(P-q)^2-M_i^2+i\epsilon}\frac{1}{q^2-m_i^2+i\epsilon},
\end{equation}
where $M_{i}$ and $m_{i}$ are the masses of baryon and meson of $i$-th coupled channel, respectively. $P$ is the four-momentum of the meson-baryon system, and $q$ is the four-momentum of meson in the center-of-mass frame. In this work, we take the dimensional regularization method, and the loop function could be written as, 
\begin{equation}\label{Eq:G-DR}
	\begin{aligned}
		G_{l}  =&i\int\frac{d^4q}{(2\pi)^4}\frac{2M_i}{(P-q)^2-M_i^2+i\epsilon}\frac{1}{q^2-m_i^2+i\epsilon} \\
		 =&\frac{2M_l}{16 \pi^2}\left\{a_l(\mu)+\ln \frac{M_l^2}{\mu^2}+\frac{s+m_l^2-M_l^2}{2 s} \ln \frac{m_l^2}{M_l^2}\right. \\
		& +\frac{|\vec{q}\,|}{\sqrt{s}}\left[\ln \left(s-\left(M_l^2-m_l^2\right)+2 |\vec{q}\,| \sqrt{s}\right)\right. \\
		& +\ln \left(s+\left(M_l^2-m_l^2\right)+2 |\vec{q}\,| \sqrt{s}\right) \\
		& -\ln \left(-s+\left(M_l^2-m_l^2\right)+2 |\vec{q}\,| \sqrt{s}\right) \\
		& \left.\left.-\ln \left(-s-\left(M_l^2-m_l^2\right)+2 |\vec{q}\,| \sqrt{s}\right)\right]\right\}.
	\end{aligned}
\end{equation}

Here, we take the regularization scale $\mu=630$ MeV and the subtraction constants to be $a_{\bar{K} N}=-1.84{\beta_1}$, $a_{\pi \Sigma}=-2{\beta_2}$ and $a_{\pi \Lambda}=-1.83{\beta_3}$~\cite{Roca:2013cca}, where $\beta_i$ are shown in Table.~{\ref{tab:1}}.

\begin{table}[htpb]
	\begin{center}	
		\caption{\label{tab:1}Parameters of the unitarized amplitudes~\cite{Roca:2013cca}.}
		\begin{tabular}{cccccccc}
			\hline\hline
		   parameters & $\alpha^1_{11}$ & $\alpha^1_{12}$ & $\alpha^1_{13}$ & $\alpha^1_{22}$ & $\beta_1$ &$\beta_2$ & $\beta_3$ \\
			\hline
			values & $0.85$   & $0.93$& $1.056$ & 0.77 & 1.187 & 0.722 & 1.119  \\
            
			\hline\hline
		\end{tabular}
	\end{center}
\end{table}

\subsection{Contributions from $N(1535)$ and $K^*(892)$}\label{Subsec:B}

In addition to the final state interaction of the $M^\prime B^\prime \to \pi^+\Lambda$, as depicted in Fig.~\ref{fig:hadron_tree_loop}(b), we must also consider the tree diagram of Fig.~\ref{fig:hadron_tree_loop}(a), and  the final state interaction of the $\pi^-p\to\Lambda K^0$, $\pi^0n\to\Lambda K^0$, $\eta n\to\Lambda K^0$, $\Lambda K^0\to\Lambda K^0$ interaction of Fig.~\ref{fig:hadron_tree_loop}(c), which can dynamically generate the intermediate state $N(1535)$. The amplitude for Figs.~\ref{fig:hadron_tree_loop}(a) and \ref{fig:hadron_tree_loop}(c) could be expressed as,

\begin{align}
\mathcal{T}^{\text{Tree}} &= V_P \left( h_{K^0\Lambda} + h_{\pi^+K^0} \right), \\
\mathcal{T}^{N(1535)} &= V_P \biggl( \sum_i h_i \tilde{G}_i t_{i\to K^0\Lambda} \notag \\
                        &\quad + h_{\pi^+ K^0} \tilde{G}_{K^0\Lambda} t_{K^0\Lambda\to K^0\Lambda} \notag \\
                        &\quad +\frac{1}{C}h_{\pi^-\pi^+}\tilde{G}_{\pi^-p} t_{\pi^-p\to K^0\Lambda}\biggr), 
\label{eq:TMB}  
\end{align}

where $i=1, 2, 3, 4$ correspond to the $\pi^-p$, $\pi^0n$, $\eta n$, and $\Lambda K^0$ channel, respectively. 
Accordingly, the coefficients $h_i$, $h_{\pi^+ K^0}$, and $h_{\pi^- \pi^+}$ can be obtained from Eqs.~(\ref{eq:MB}), (\ref{eq:7b}), and (\ref{eq:9d}), respectively, and the values of $h_i$ are given by

\begin{align}
&h_{\pi^- p}=\frac{1}{\sqrt{2}},~h_{\pi^0 n}=-\frac{1}{2},~h_{\eta n}=\frac{1}{\sqrt{6}}, \notag \\
&h_{K^0 \Lambda}=\frac{1}{\sqrt{3}},~h_{\pi^+ K^0}=\frac{1}{\sqrt{3}},~h_{\pi^- \pi^+}=\frac{1}{\sqrt{2}}. 
\label{eq:hiN1535}
\end{align}


The transition amplitude of the coupled channels are $t_{i \to K^0\Lambda}$, which is obtained by solving the Bethe-Salpeter equation as Eq.~(\ref{Eq:BS}), where the transition potential $V_{ij}$ is taken from Refs.~\cite{Wang:2015pcn,Inoue:2001ip},
\begin{eqnarray}
\begin{aligned}
V_{ij}& =-\frac{C_{ij}}{4f_if_j}\left(2\sqrt{s}-M_i-M_j\right)  \\
&\times\left(\frac{M_i+E_i}{2M_i}\right)^{1/2}\left(\frac{M_j+E_j}{2M_j}\right)^{1/2},
 \label{eq:Vij}
\end{aligned}
 \end{eqnarray}
where $E_i$ and $M_i$ are the energy and mass of the baryon in the $i$-th channel, and the coefficients $C_{ij}$ reflecting the SU(3) flavor symmetry are obtained by Ref.~\cite{Lyu:2023aqn}.  The coupling constant $f$ is given by,
 \begin{eqnarray}
f_{\pi}=93\mathrm{~MeV},\quad f_{K}=1.22f_{\pi},\quad f_{\eta}=1.3f_{\pi}.
 \end{eqnarray}

In this part, the loop function $\tilde{G}_i$ of the meson-baryon system in Eq.~(\ref{eq:TMB}) is calculated by using cutoff method,
  \begin{eqnarray}
 \begin{aligned}
\tilde{G}_{l}& \begin{aligned}&=i\int\frac{d^4q}{(2\pi)^4}\frac{2M_l}{(p-q)^2-M_l^2+i\epsilon}\frac{1}{q^2-m_l^2+i\epsilon}\end{aligned}  \\
&&\begin{matrix}\\\end{matrix} \\
&=\int_0^{q_{\max}}\frac{2M_l}{(2 \pi)^2}\cdot \dfrac{|q|^2(\omega_1+\omega_2)dq}{\omega_1\omega_2[s-(\omega_1+\omega_2)^2]},
\end{aligned}
 \label{eq:G-cutoff}
 \end{eqnarray}
with $\omega=\sqrt{m^{2}+\vec{q}^{\,2}}$. And the cutoff momentum is taken with $q_{\max}=1200$ MeV. However, the loop function $G$ in $t_{i \to K^0\Lambda}$ is given by the dimensional regularization method, and we take the regularization scale $\mu=1200$ MeV and use the following values for the subtraction constants $a_i$~\cite{Wang:2015pcn,Inoue:2001ip,Lyu:2023aqn},
\begin{equation}
	\begin{aligned}
		&a_{K^+\Sigma^-}=-2.8,~a_{K^0\Sigma^0}=-2.8,~a_{K^0\Lambda}=1.6,\\
		&a_{\pi^-p}=2.0,~~~~~~a_{\pi^0n}=2.0,~~~~~~a_{\eta n}=0.2.
	\end{aligned} \label{eq:subc}
\end{equation}

It should be noted that, in a single project, it would be better to use the same regularization method to calculate these loop functions. But one can find that, in Eq.~(\ref{eq:subc}), the subtraction constants $a_i(\mu)$ for the channels of $K^0\Lambda$, $\eta n$, $\pi^-p$ and $\pi^0 n$ are positive. With a cutoff $q_{\rm max}$ in the cutoff method, the matrix $G_i$ of Eq.~(\ref{Eq:G-DR}) (used by Eq.~(\ref{Eq:BS})) would imply negative subtraction constants $a_i(\mu)$, not positive ones. The need for values $a_i(\mu)>0$ is an indication that one is including the contribution of missing channels in the scattering amplitude~\cite{Hyodo:2008xr}. However, the primary $\Lambda_c^+\rightarrow \pi^+ MB$ is selective to just four channels, with particular weights, which then propagate by means of the $\tilde{G}_i$ in Eq.~(\ref{eq:TMB}) function. We are not justified to use the $\tilde{G}_i$ function of regularization method in Eq.~(\ref{eq:TMB}) of scattering to account for channels which would not contribute there~\cite{Hyodo:2008xr}. Therefore, to evaluate this amplitude, we employ dimensional regularization for the function $G$ in Eq.~(\ref{Eq:BS}) which appears in the $t_{i,K^0\Lambda}$ term of Eq.~(\ref{eq:TMB}) and the cutoff method for $\tilde{G}$ in Eq.~(\ref{eq:TMB}).

\begin{figure}[htbp]
\begin{center}
\includegraphics[width=0.45\textwidth]{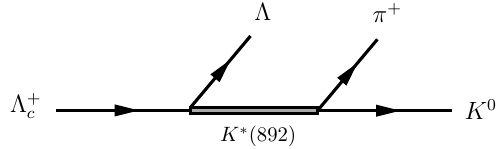}
\caption{Mechanism for the $\Lambda^+_c \to \Lambda K^0 \pi^+$ decay via the intermediate $K^*(892)$.} 
\label{fig:K892}
\end{center}
\end{figure}

On the other hand, in addition to the mechanisms of Fig.~\ref{fig:hadron_tree_loop}, the process $\Lambda^+_c \to \Lambda K^0 \pi^+$ could also happen through the mechanism of $\Lambda^+_c \to \Lambda K^*(892) \to \Lambda K^0 \pi^+$, as depicted in Fig.~\ref{fig:K892}. Then, we write the decay amplitude, as follows,

\begin{equation}\label{eq:T-K}
\mathcal{T}^{K^*}=V_{P}^{\prime}\frac{\vec{\epsilon}_i\vec{p}_{\Lambda}^{~i}\vec{\epsilon}_j\vec{p}_{\pi}^{~j}}{M_{\pi^+K^0}^2-M_{K^*}^2+iM_{K^*}\Gamma_{K^*}}.
\end{equation}
After summing over polarizations of the $\epsilon$, we have
\begin{equation}
\mathcal{T}^{K^*}=V_{P}^{\prime}\frac{|\vec{p}_{\pi^+}||\vec{p}_\Lambda|\cos\theta}{M_{\pi^+K^0}^2-M_{K^*}^2+iM_{K^*}\Gamma_{K^*}},
\end{equation}
where $V_{P}^{\prime}$ is the relative strength of the contribution from the intermediate resonance $K^*$, and $|\vec{p}_{\pi^+}|$ and $|\vec{p}_\Lambda|$ are the momenta of $\pi^+$ and $\Lambda$ in the rest frame of the $\pi^+K^0$ system, respectively, and $\theta$ is the angle between $\pi^+$ and $\Lambda$ in the center-of-mass frame of the $\pi^+K^0$ system, which can be given by~\cite{Lyu:2024wxa,Lyu:2024qgc}
\begin{equation}
\begin{aligned}
|\vec{p}_{\pi^+}| & =\frac{\lambda^{1/2}(M_{\pi^+K^0}^2,m_{\pi^+}^2,m_{K^0}^2)}{2M_{\pi^+K^0}}, \\
\left|\vec{p}_\Lambda\right| & =\frac{\lambda^{1/2}(M_{\Lambda_c^+}^2,m_\Lambda^2,M_{\pi^+K^0}^2)}{2M_{\pi^+K^0}},  \\
\cos\theta & =\frac{M_{\Lambda K^0}^2-M_{\Lambda_c^+}^2-m_{\pi^+}^2+2P_{\Lambda_c^+}^0P_{\pi^+}^0}{2|\vec{p}_{\pi^+}||\vec{p}_\Lambda|},
\end{aligned}
\end{equation}
where $P_{\Lambda_c^+}^{0}$, $P_{\pi^+}^{0}$ are the $\Lambda_c^+$ and $\pi^+$ energies in the $\pi^+K^0$ rest frame as follows
\begin{equation}
\begin{aligned}
P_{\Lambda_c^+}^{0} &= \sqrt{M_{\Lambda_c^+}^{2}+|\vec{p}_{\Lambda}|^{2}}, \\
P_{\pi^+}^{0} &= \sqrt{m_{\pi^+}^{2}+|\vec{p}_{\pi^+}|^{2}}.
\end{aligned}
\end{equation}

The values $M_{K^*}=891.67$~MeV and $\Gamma_{K^*}=51.4$~MeV are used in our calculation.


\subsection{Invariant mass distributions} \label{Subsec:C}

Using the formalism mentioned above, we can detail the modulus squared of the total decay amplitude of the process $\Lambda^+_c \to \Lambda K^0 \pi^+$ as follows:
  \begin{align}
|\mathcal{T}|^{2}= |\mathcal{T}^{Tree} + \mathcal{T}^{{N(1535)}}+\mathcal{T}^{{\Sigma^*(1/2^-)}}e^{i \phi}+\mathcal{T}^{{K^*}}e^{i \phi^\prime}|^{2},
\label{Eq:T-total3}
\end{align} 
where $\phi$ and $\phi^\prime$ are the relative phase angles of $\Sigma^*(1/2^-)$ and $K^*(892)$, respectively. Then the double differential width of the process $\Lambda^+_c \to \Lambda K^0 \pi^+$ is given as follows:
\begin{eqnarray}
\frac{d^{2}\Gamma}{dM_{K^{0}\Lambda}dM_{\pi^{+}\Lambda}}=\frac{1}{(2\pi)^{3}}\frac{M_{\Lambda} M_{K^{0}\Lambda}M_{\pi^{+}\Lambda} }{2M_{\Lambda_{c}^{+}}^{2}}|\mathcal{T}|^{2},
\end{eqnarray}
with the Mandl and Shaw normalization of the meson and baryon fields~\cite{Mandl:1985bg}.

For a given value of $M_{12}$, the range of $M_{23}$ is determined according to the RPP~\cite{ParticleDataGroup:2024cfk},
\begin{align}
	&\left(m_{23}^2\right)_{\min}=\left(E_2^*+E_3^*\right)^2-\left(\sqrt{E_2^{* 2}-m_2^2}+\sqrt{E_3^{* 2}-m_3^2}\right)^2, \nonumber\\
	&\left(m_{23}^2\right)_{\max}=\left(E_2^*+E_3^*\right)^2-\left(\sqrt{E_2^{* 2}-m_2^2}-\sqrt{E_3^{* 2}-m_3^2}\right)^2, \label{eq:limit}
\end{align}
where $E_{2}^*$ and $E_{3}^*$ are the energies of particles 2 and 3 in the $M_{12}$ rest frame, which are written as
\begin{eqnarray}
E_{2}^{*}=\frac{M_{12}^2-m_1^2+m_2^2}{2M_{12}},\\E_{3}^{*}=\frac{M_{\Lambda_{c}^{+}}^{2}-M_{12}^{2}-m_{3}^{2}}{2M_{12}},
\label{eq:e23}
\end{eqnarray}
while $m_1$, $m_2$, and $m_3$ are the masses of particles 1, 2, and 3, respectively. Permutation of the indices allows us to evaluate all three mass distributions, using $M_{12}$, $M_{23}$ as independent variables, and the property $M_{12}^2+M_{13}^2+M_{23}^2=M_{\Lambda_c^+}^2+m_{\pi^+}^2+m_{K^0}^2+M_\Lambda^2$ to get $M_{13}$ from them. The masses and widths of the particles are sourced from the RPP~\cite{ParticleDataGroup:2024cfk}. 

Furthermore, one can see that, in above formalism, we have two parameters $V_P$ and $V_{P}^{\prime}$, which represent the relative strength from the mechanisms of Fig.~\ref{fig:hadron_tree_loop}, and Fig.~\ref{fig:K892}, respectively. In this work, we can calculate the values utilizing the branching fraction $\mathcal{B}(\Lambda^+_c \to \Lambda K^0 \pi^+)=2\times \mathcal{B}(\Lambda^+_c \to \Lambda K^0_S \pi^+)= 2\times 1.73\times 10^{-3}$~\cite{BESIII:2024xny}, and $\mathcal{B}(\Lambda^+_c \to \Lambda K^{*}\to\Lambda K^0 \pi^+)= 2/3\times \mathcal{B}(\Lambda^+_c \to \Lambda K^{*})=  2/3\times2.4\times 10^{-3}$~\cite{BESIII:2024xny}. The values of the $V_P$ and $V_{P}^{\prime}$ are listed in Table.~\ref{tab:vp}

\begin{table}[htpb]
	\begin{center}	
		\caption{Values of the parameters  $V_P$ and $V_{P}^{\prime}$.}
		\begin{tabular}{ccc}
			\hline\hline
		   Parameters \qquad\quad&  $V_P~(\text{MeV}^{-1})$ \qquad\quad&$V_{P}^{\prime}~(\text{MeV}^{-1})$ \\
			\hline
			Values \qquad\quad& $1.60 \times 10^{-8}$   \qquad\quad& $3.07 \times 10^{-9}$  \\
            
			\hline\hline
		\end{tabular}\label{tab:vp}
	\end{center}
\end{table}

Although several other $N^*$ resonances exist in the $\Lambda K^0$ invariant mass distribution from 1600 to 2150~MeV, including these intermediate states would increase the number of free parameters. Furthermore, considering that these states do not affect the structure of $\Sigma(1/2^-)$, we have omitted their contributions. Future measurements of this process could provide more information about these excited $N^*$ states.

\section{Numerical results and discussion} \label{Sec:Results}

\begin{figure}[htbp]
\centering
\subfigure
{
        \centering
        \includegraphics[scale=0.55]{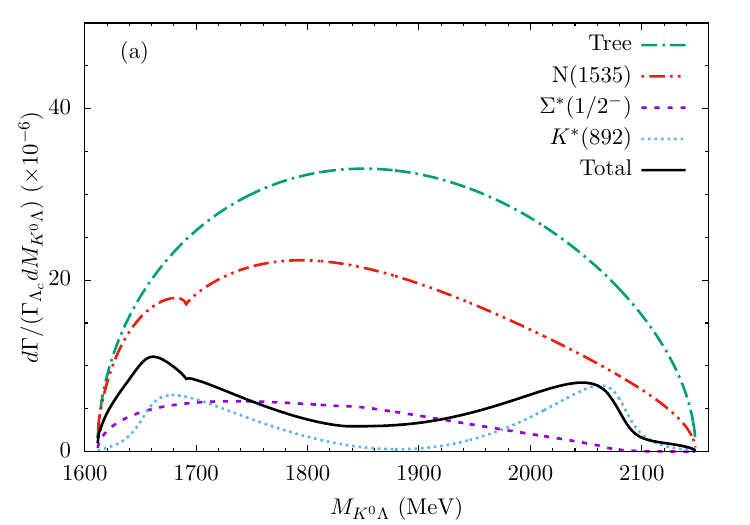}
}
\subfigure
{
        \centering
        \includegraphics[scale=0.55]{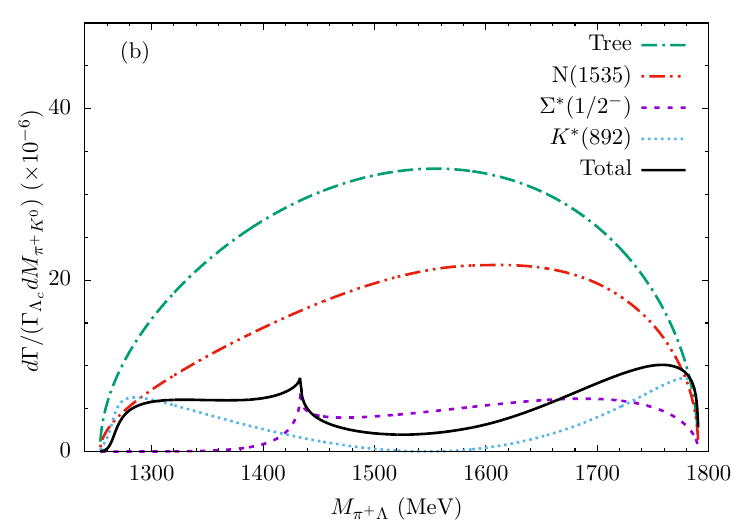}
}
\subfigure
{
        \centering
        \includegraphics[scale=0.55]{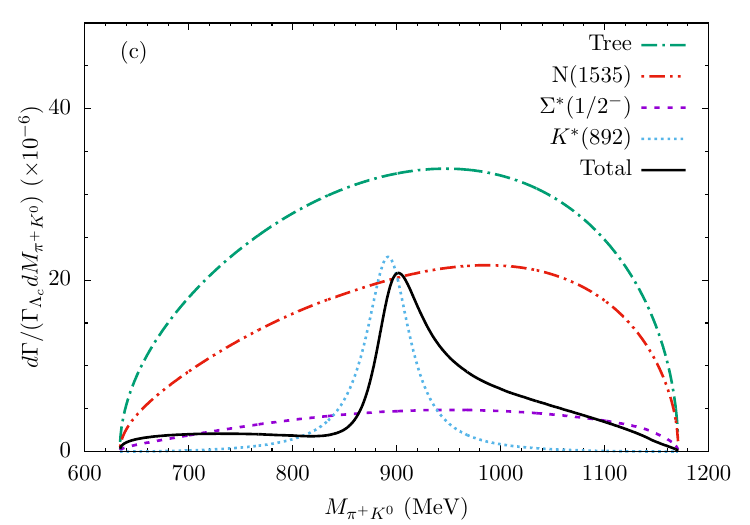}
}
\caption{The $K^0\Lambda$ (a), $\pi^+\Lambda$ (b) and $\pi^+K^0$ (c) differential branching fractions of the $\Lambda_c^+ \to \Lambda K^0 \pi^+$ decay with the phase angles $\phi=\phi'=0$.}
\label{fig:no_para}
\end{figure}

            


\begin{figure}[htbp]
\centering
\subfigure
{
        \centering
        \includegraphics[scale=0.55]{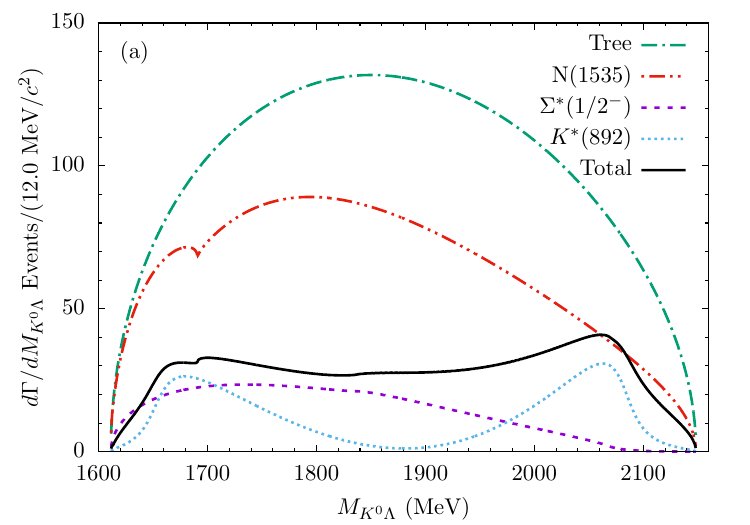}
}
\subfigure
{
        \centering
        \includegraphics[scale=0.55]{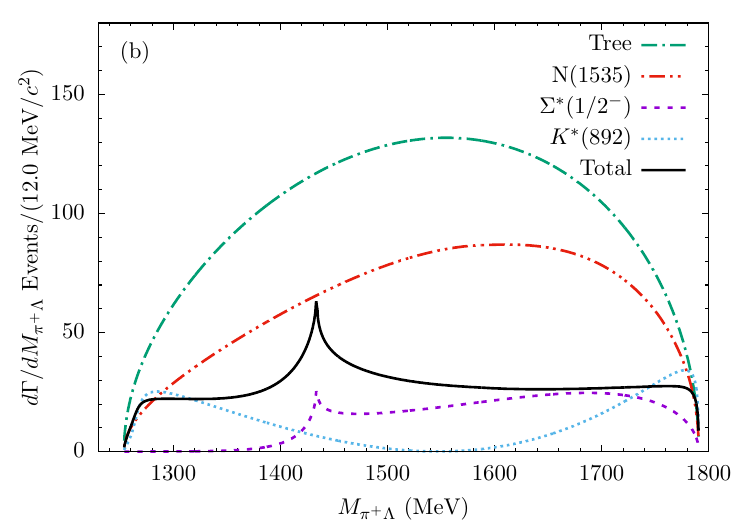}
}
\subfigure
{
        \centering
        \includegraphics[scale=0.55]{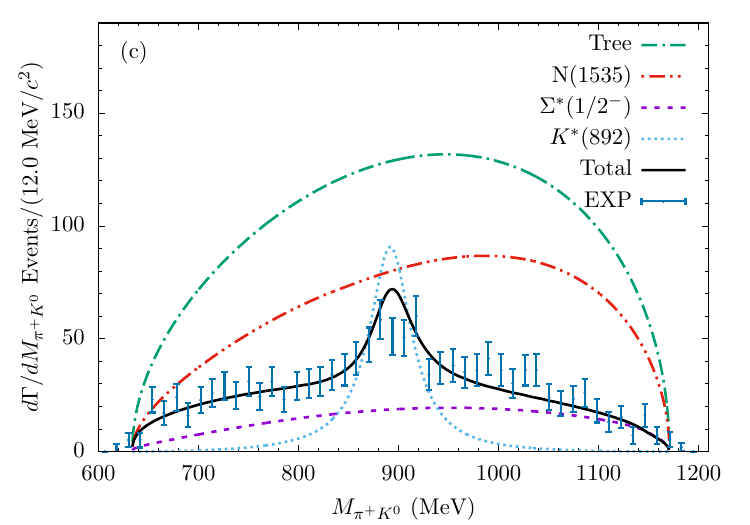}
}
\caption{The $K^0 \Lambda $(a), $\pi^+ \Lambda $(b) and $\pi^+ K^0 $(c) invariant mass distributions of the process $\Lambda^+_c \to \Lambda K^0 \pi^+$. The experimental data labeled by ``EXP" are obtained from  Ref~\cite{BESIII:2024xny}.}
\label{fig:three_para}
\end{figure}


\begin{figure}[htbp]
    \centering
    \subfigure[]{\includegraphics[width=0.44\textwidth]{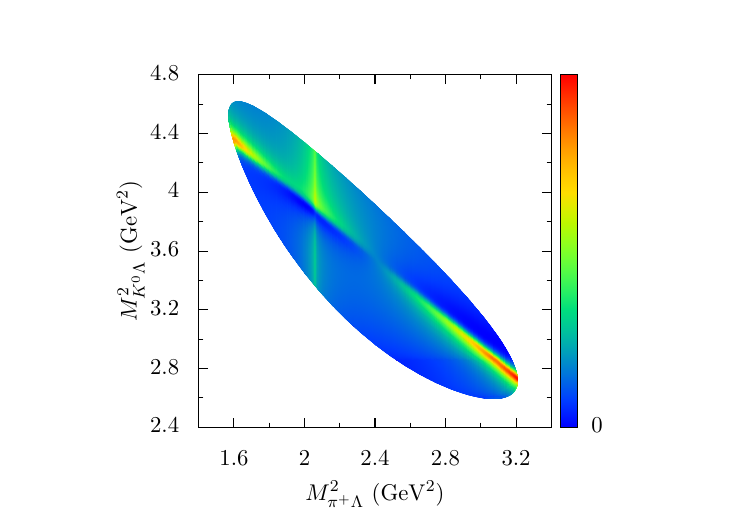}}
    \hfill  
    \subfigure[]{\includegraphics[width=0.44\textwidth]{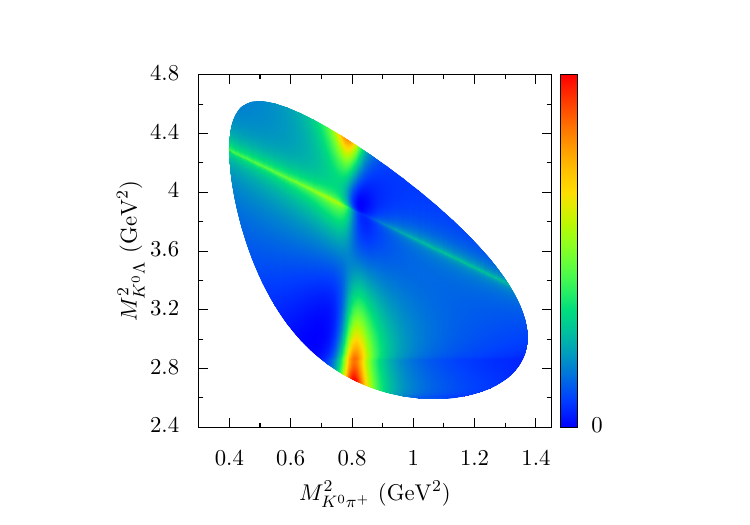}}
    \hfill
    \subfigure[]{\includegraphics[width=0.44\textwidth]{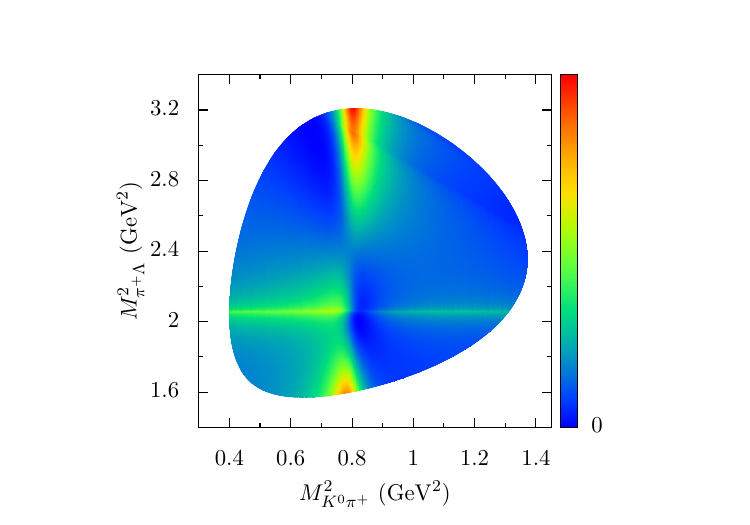}}
    
    \caption{Dalitz plots of $M_{\pi^+ \Lambda}^2$ vs $M_{K^0 \Lambda}^2$ (a), $M_{K^0 \pi^+}^2$ vs $M_{K^0 \Lambda}^2$ (b), and $M_{K^0 \pi^+}^2$ vs $M_{\pi^+ \Lambda}^2$ (c) for the process $\Lambda_c^+ \to \Lambda K^0 \pi^+$, with fitted parameters.}
    \label{fig:dalitz}
\end{figure}


Firstly, we calculated the invariant mass distributions for the process $\Lambda^+_c \to \Lambda K^0 \pi^+$ taking the phase angles $\phi=\phi'=0$, where, in this case, there is no free parameter. Figure~\ref{fig:no_para} shows the $K^0 \Lambda $, $\pi^+ \Lambda $, and $\pi^+ K^0 $ invariant mass distributions of the process $\Lambda^+_c \to \Lambda K^0 \pi^+$. The green-dot-dashed curves represent the contribution from the tree diagram, the red-dashed-dotted curves show the contribution from the $N(1535)$ state, the purple-dashed curves show the contribution from $\Sigma^*(1/2^-)$, the blue-dotted curves show the contribution from $K^*(892)$, and the black-solid curves show the total results.
One can see a threshold enhancement structure in $\Lambda K^0$ invariant mass distribution, which is due to the nucleon resonance $N(1535)$~\cite{Liu:2005pm}. On the other hand, one can find a clear cusp structure around 1430~MeV in $\Lambda\pi^+$ invariant mass distribution, which is associated with $\Sigma^*(1/2^-)$. Furthermore, in $\pi^+K^0$ invariant mass distribution, a significant peak is observed around 890~MeV, which can be attributed to intermediate $K^*(892)$.



As mentioned in the introduction, the BESIII Collaboration reported the $\pi^+K^0_S$ invariant mass distribution of signal events for the process $\Lambda^+_c \to \Lambda K^0_S \pi^+$.  In order to compare our theoretical results with the BESIII measurements of the $\pi^+K^0_S$ invariant mass distribution, it is necessary to introduce a normalization constant. Therefore, in addition to the normalization constant, our model contains two free parameters: the phase angles $\phi$ and $\phi'$ appeared in Eq.~(\ref{Eq:T-total3}). The fitting yields $\phi=(1.73\pm 0.11)\pi$, $\phi^\prime=(1.57\pm 0.06)\pi$, and a normalization constant $(1.22 \pm 0.26)\times10^{15}$, with $\chi^2/\text{d.o.f}=0.98$, and the corresponding invariant mass distributions calculated with these fitted parameters are presented in Fig.~\ref{fig:three_para}. 
It can be observed that our calculated $\pi^+K^0$ invariant mass distribution are in good agreement with the BESIII measurements~\cite{BESIII:2024xny}. Moreover, the cusp structure associated with the $\Sigma^*(1/2^-)$ remains clearly visible in the $\pi^+\Lambda$ invariant mass distribution.


In addition, as shown in Fig.~\ref{fig:dalitz}, with the fitted parameters we present the Dalitz plots of ``$M^2_{\pi^+ \Lambda}$" vs ``$M^2_{K^0 \Lambda}$", ``$M^2_{K^0 \pi^+}$" vs ``$M^2_{K^0 \Lambda}$" and ``$M^2_{K^0 \pi^+}$" vs ``$M^2_{\pi^+ \Lambda}$" for the process $\Lambda^+_c \to \Lambda K^0 \pi^+$. One can clearly find the signals of the resonances $\Sigma^*(1/2^-)$ and $K^*(892)$, while no clear structure is found for $N(1535)$ in the Dalitz plot because it lies below the threshold.

In all, these results could be tested by BESIII, Belle~II, and the proposed Super Tau-Charm Factory~\cite{Guo:2022kdi,Cheng:2022tog} experiments in the future. The more precise measurement is crucial to finding the evidence of $\Sigma^*(1/2^-)$ and understanding its role in the process $\Lambda_c^+\to\Lambda K^0\pi^+$.

\section{Summary}
\label{Sec:summary}

Recently, the BESIII Collaboration has analysed the singly Cabibbo-suppressed decays $\Lambda^+_c \to \Lambda K^+ \pi^0$ and $\Lambda^+_c \to \Lambda K_s^0 \pi^0$, and obtained the branching fractions $\mathcal{B}(\Lambda_c \to \Lambda\pi^+K^0) = (1.5\pm0.3)\times10^{-3}$ and $\mathcal{B}(\Lambda_{c}^{+} \to \Lambda K_S^{0}\pi^{+}) = (1.73 \pm 0.27 \pm 0.10)\times 10^{-3}$, respectively, which 
provides an important lab to search for the predicted $\Sigma^*(1/2^-)$ resonance.

In this work, we analyzed the process $\Lambda^+_c \to \Lambda K^0\pi^+$, by considering the intermediate states $K^*(892)$ and the $S$-wave final state interaction within the chiral unitary approach, which could dynamically generate the $\Sigma^*(1/2^-)$ and the $N(1535)$.
By fitting to the BESIII measurements of the $\pi^+K^0$ invariant mass distribution, we obtained the $\chi^2/\text{d.o.f}=0.98$, and our results could well reproduce the BESIII data. Meanwhile, one can find  a threshold enhancement structure in the $K^0\Lambda$ invariant mass distribution, associated with $N(1535)$, and  a significant cusp structure in the $\pi^+\Lambda$ invariant mass distribution, associated with the $\Sigma^*(1/2^-)$.
Future high-precision measurements at BESIII, Belle II, and the proposed Super Tau-Charm Facility (STCF) are essential to test these predictions.

In summary, the predictions of $\Sigma^*(1/2^-)$ signal in $\Lambda\pi^+$ invariant mass distribution could be tested by BESIII, Belle~II, and the proposed Super Tau-Charm Factory~\cite{Guo:2022kdi,Cheng:2022tog} experiments in the future. Experimental confirmation would not only establish a new low-lying baryon resonance but also advance our understanding of QCD in the non-perturbative regime, ultimately contributing to a more complete picture of the hadron spectrum.




\section{ACKNOWLEDGMENTS}

E.Wang acknowledge the support from the National Key R\&D Program of China (No. 2024YFE0105200).
This work is supported by the Natural Science Foundation of Henan under Grant No. 252300423951, the National Natural Science Foundation of China under Grant No. 12205075, No. 12475086, No. 12192263 and No. 12335006. This work is also supported by Zhengzhou University Young Student Basic Research Projects (PhD students) under Grant No. ZDBJ202522.

\bibliographystyle{ursrt}

\end{document}